\documentclass[12pt]{article}
\usepackage{graphicx}
\def\be{\begin{equation}}
\def\ee{\end{equation}}
\def\bi{\bibitem}
\def\ci{\cite}
\def\mc{\mathcal}
\def\gn{\Gamma_n}

\begin{document}

\begin{center}
\bf{\large Complex masses of resonances and the Cornell potential}\\
\vspace{3mm}
\rm{ M. N. Sergeenko }\\
\vspace{2mm}
{\small\it{Stepanov Institute of Physics, 
Belarus National Academy of Sciences\\ 
68 Nezavisimosti Ave., 220072 Minsk, Belarus}}\\
{\small\ E-mail: msergeen@usa.com}
\end{center}

\begin{abstract}
Physical properties of the Cornell potential in the complex-mass 
scheme are investigated. Two exact asymptotic solutions of relativistic 
wave equation for the coulombic and linear components of the potential 
are used to derive the resonance complex-mass formula. The centered 
masses and total widths of the $\rho$-family resonances are calculated. 
\vskip 5mm
\noindent Pacs: 11.10.St; 12.39.Pn; 12.40.Nn; 12.40.Yx 

\end{abstract}

\centerline{\bf 1. Introduction}
\vskip 3mm\noindent 
Resonance is the tendency of a system to oscillate at a greater 
amplitude at some frequencies. Most particles listed in the 
Particle Data Group tables \ci{PDG2012} are unstable. A thorough 
understanding of the physics summarized by the PDG is related to the 
concept of a resonance. 

There are great amount and variety of experimental data and the 
different approaches used to extract the intrinsic properties of the 
resonances. In Quantum Mechanics and Quantum Field Theory resonances 
may appear in similar circumstances to classical physics. They can also 
be thought of as unstable particles with the particle's complex energy 
poles in the scattering amplitude.

As known, operators in QM are Hermitian and the corresponding 
eigenvalues are real. However, in scattering experiment, the wave 
function requires another boundary condition (different from bound 
state's), that is why the complex energy is required 
\ci{NiMoisey,Taylor}. 

It was in 1939 that Siegert introduced the concept of a purely outgoing 
wave belonging to the complex eigenvalue $\mc E_n=E_n-i\gn/2$ 
as an appropriate tool in the studying of resonances \ci{Sieg}. 
This complex eigenvalue also corresponds to a first-order pole of 
the S-matrix \ci{Heit}.

In elastic and inelastic scattering experiments the resonances are 
associated with the scattering particle and the target \ci{RuppCB}. 
In such case the energy decreases exponentially with the time so that 
the damping constant $\Gamma$ is a measure of the lifetime, 
$\tau=h/\Gamma$, of the oscillation. 

Resonances in QFT are described by the 
complex-mass poles of the scattering matrix \ci{Taylor}. Resonance 
is present as {\em transient oscillations} associated with metastable 
states of a system which has sufficient energy to break up into two 
or more subsystems. The masses of intermediate particles develop 
imaginary masses from loop corrections \ci{BauDGST}. In this case, 
the probability density comes from the particle's propagator, with 
the complex mass, 
\be
\mc M=M-i\Gamma/2. \label{Mcx}\ee
This formula is related to the particle's decay rate by the optical 
theorem \ci{Taylor,LandLif}. 

The hadronic resonances are never observed directly but through their 
decay channels; this causes problems with their definition. There is 
the lack of a precise definition of what is meant by mass and width 
of resonance. Comprehensive definitions of the centered mass and 
width of resonances require further investigations \ci{BerniCaPe}. 

Fundamentals of scattering theory and strict mathematical definition 
of resonances in QM was considered in \ci{Hislop}. 
The rigorous QM definition of a resonance requires determining the 
pole position in the second Riemann sheet of the analytically 
continued partial-wave scattering amplitude in the complex 
Mandelstam $s$ variable plane \ci{NieArri}. 

Resonances are defined as the poles of the meromorphic continuation of 
the cut-off resolvent, which are shown to be the same as the poles of 
the meromorphically continued S-matrix. This definition has the 
advantage of being quite universal regarding the pole position, but 
can only be applied if the amplitude can be analytically continued in 
a reliable way. 

In particle physics resonances arise as unstable intermediate states 
with complex masses. The advantage of analyzing a system in the complex 
plane has important features such as a simpler and more general 
framework. Complex numbers allow to get more than what we insert. 
The complex-mass scheme provides a consistent framework for dealing 
with unstable particles and has been successfully applied to various 
loop calculations \ci{BauDGST}. 

In traditional approach to investigate resonances one deals with 
the scattering theory, exploring the properties of S-matrix and 
partial amplitudes. In this work, in contrast to the usual analysis, 
we consider mesonic resonances to be the quasi-bound states of a 
hadron constituents. We solve the bound-state problem in the potential 
approach and analyze the mass spectrum generated from solution of 
the relativistic wave equation for the Cornell potential. We show, 
that two asymptotic components of the potential, the coulombic term 
and linear one yield the complex masses of resonances. This results in 
the complex-mass formula for the resonances that allows to calculate 
in unified way their centered masses and total widths.\\

\centerline{\bf 2. The universal mass formula}
\vskip 3mm\noindent
In perturbative QCD, as in QED the essential interaction at small 
distances is instantaneous coulombic one-gluon exchange (OGE); in QCD, 
it is $qq$, $qg$, or $gg$ Coulomb scattering \ci{Bjor}. Therefore, one 
expects from OGE a Coulomb-like contribution to the potential, i.e., 
$V_S(r)\propto-\alpha_s/r$ at $r\rightarrow 0$.

For large distances, in order to be able to describe confinement, 
the potential has to rise to infinity. From lattice-gauge-theory 
computations \ci{LattV} follows that this rise is an approximately 
linear, i.e., $V_L(r)\simeq\sigma r+$const for large $r$, where
$\sigma\simeq 0.15$\,GeV$^2$ is the string tension. These two 
contributions by simple summation lead to the famous Cornell 
$q\bar q$ potential \ci{LattV,Eich},
\be
V(r)=V_S(r)+V_L(r)\equiv-\frac 43\frac{\alpha_s}r +\sigma r;
\label{Vcor}\ee
its parameters are directly related to basic physical quantities 
noted above. All phenomenologically acceptable QCD-inspired 
potentials are only variations around this potential.

This potential is one of the most popular in hadron physics and 
incorporates in clear form the basic features of the strong 
interaction. In hadron physics, the nature of the potential is very 
important. There are normalizable solutions for scalarlike potentials, 
but not for vectorlike. The effective interaction has to be 
Lorentz-scalar in order to confine quarks and gluons \ci{Su}.
In our consideration, we take the potential (\ref{Vcor}) to be 
Lorentz-scalar.  

It is hard to find an analytic solution of known relativistic wave 
equations for the potential (\ref{Vcor}). This aim can be achieved 
with the use of the semi-classical wave equation \ci{SeMPL}. 
An important feature of this equation is that, for two and more 
turning-point problems, it can be solved exactly by the 
conventional WKB method \ci{SeMPL,SePRA}.

Our aim is to find in analytic form the energy/mass eigenvalues 
for the potential (\ref{Vcor}). This is not easy task even for 
the quasiclassical method. This is why, using the two-point Pad\'e 
approximant, we joined two exact solutions obtained separately 
for the short-distance coulombic part, $V_S(r)$, and long-distance 
one, $V_L(r)$, of the Cornell potential. As a result we obtained 
the interpolating mass formula \ci{SeEPJC12,SeEPL10,SeYF93,SeZC94},
\be
M_n^2=4\left[2\sigma\tilde N 
-\left(\frac{\tilde\alpha m}N\right)^2
+m^2-2\tilde\alpha\sigma\right],
\label{Mn2int}\ee 
where $\tilde\alpha=(4/3)\alpha_s$, $\tilde N=N+n_r+1/2$, $N=n_r+J+1$ 
and $m$ is the constituent quark mass. 
The simple mass formula (\ref{Mn2int}) describes equally well the 
mass spectra of all $q\bar q$ and $Q\bar Q$ mesons ranging from the 
$u\bar d$ ($d\bar d$, $u\bar u$, $s\bar s$) states up to the heaviest 
known $b\bar b$ systems \ci{SeYF93,SeZC94}. 

The obtained from Eq. (\ref{Mn2int}) ``saturating'' Regge trajectories 
\ci{SeYF93,SeZC94} were applied with success to Compton scattering, 
vector meson photoproduction and the partonic structure of the nucleon 
\ci{CaLagPRD02}, the photoproduction of vector mesons that provide an 
excellent simultaneous description of the high and low $-t$ behavior 
of the $\gamma\,p\rightarrow p\,\rho$, $\omega$, $\phi$ cross sections 
\ci{CLAS_PRL03}, to interpretation of the space-time structure of hard 
scattering processes \ci{LagetPRD04}, given an 
appropriate choice of the relevant coupling constants (JML-model) 
\ci{CLAS_PRL01,CLAS_EPJA05}. It was shown that the hard-scattering 
mechanism is incorporated in an effective way by using the 
``saturated'' Regge trajectories that are independent of $t$ at 
large momentum transfers \ci{SeZC94,SeYF93,PatRossi03,HuberLettInt}. 

The universal formula (\ref{Mn2int}) has been used to calculate the 
glueball masses and Regge trajectories including the Pomeron 
\ci{SeEPJC12,SeEPL10}. It appears to be successful in many 
applications and can be justified with the use of the complex-mass 
scheme.\\

\centerline{\bf 3. The complex-mass eigenvalues}
\vskip 3mm\noindent
Resonances are complex values and can be described by complex numbers. 
These numbers are important even if one wants to find real solutions 
of a problem. Using complex numbers, we are getting more than what we 
insert. Remind the important properties of complex numbers such 
as the {\it fundamental theorem of algebra}, i.e., the existence of 
$n$ roots of any $n$-th order polynomial with complex coefficients. 
As known, it wouldn't work if we demanded real solutions. 

Holomorphic (natural) functions of a complex variable have many 
important mathematical properties that turn complex numbers into 
useful if not essential tools, e.g. in the case of two-dimensional 
conformal field theories (CFT). In many of the applications, the 
complex numbers may be viewed as non-essential but very useful 
technical tricks.

However, operators in QM are Hermitian and the corresponding eigenvalues 
are real. But, the complex eigenvalues are well known in physics 
\ci{NiMoisey,Sieg,Heit}. The wave function in scattering experiment  
requires another boundary condition, which results in the complex-energy 
eigenvalues. The complex eigenvalues also correspond to a first-order 
pole of the S-matrix \ci{Heit}.\\

{\it 3.1. The complex eigenmomenta}.

The mass formula (\ref{Mn2int}) is very transparent physically, as well 
as the potential (\ref{Vcor}) (OGE + linear). This formula contains in 
a hidden form some important information. We can get it in the following 
way. It is easy to see that eq. (\ref{Mn2int}) is the real-part mass 
squared of the complex equation,
\be 
\mc M_n^2=4[(\pi_n)^2+\mu^2], \label{E2nCx}\ee 
where
\be
\pi_n=\sqrt{2\sigma\tilde N}-i\frac{\tilde\alpha m}N, 
\label{PnCx}\ee
\be 
\mu=m+i\sqrt{2\tilde\alpha\sigma}. \label{MCx}\ee 
The complex-mass expression (\ref{E2nCx}) contains more important 
information, but first, let us give some ground to our consideration. 

An important hint is related to the hydrogen atom. 
The classic Coulomb problem in QM can be viewed in terms of complex 
values. One can easily observe that the {\it total} energy eigenvalues 
for the non-relativistic Coulomb problem can be written with the use 
of complex quantities in the form of kinetic energy for a free particle, 
\be
E_n=\frac{p_n^2}{2m},\ \ \ p_n=\frac{i\alpha m}{n_r+l+1}.
\label{Etokin}\ee
Here $p_n=mv_n$ is the electron's momentum eigenvalue with the 
imaginary discrete velocity, $v_n=i\alpha/(n_r+l+1)$. This means, that 
the motion of the electron in a hydrogen atom is free, but restricted 
by the ``walls'' of the potential (restricted freedom) 
\ci{SeMPL,SePRA,ClaSo}. 

Now, consider the Cornell potential (\ref{Vcor}) which is unique in that 
sense, it yields the {\it complex} masses of resonances. To show that, 
analyze the eigenvalues obtained separately for two components of the 
potential (\ref{Vcor}), i.e., the coulombic term and the linear one. 

Relativistic two-body Coulomb problem for two particles of equal masses 
can be solved analytically. The exact expression for the center-of-mass 
energy squared is well known and can be written in the form of two free 
relativistic particles as \ci{SeEPJC12,SeEPL10}:
\be
E_n^2=4[(i{\rm Im}\,\pi_n)^2+m^2],\ \ \ 
{\rm Im}\,\pi_n=-\frac{\tilde\alpha m}N,\label{En2Cou}\ee
where $N$ is given above. Here we have introduced the {\it imaginary} 
momentum eigenvalues, Im$\,\pi_n$.

The linear term of the Cornell potential (\ref{Vcor}) can be dealt 
with analogously. In this case the exact solution is also well known 
\ci{SeMPL,SeEPJC12,SeEPL10}:
\be
E_n^2=8\sigma\tilde N, \ \ \ \tilde N=N+n_r+1/2. \label{E2li}\ee
This expression does not contain the mass term and can be written 
in the form of the energy squared for two free relativistic particles,
\be
E_n^2=4({\rm Re}\,\pi_n)^2,\ \ \ 
{\rm Re}\,\pi_n =\sqrt{2\sigma\tilde N}, \label{E2Re}\ee
where ${\rm Re}\,\pi_n$ is the {\it real} momentum eigenvalue.  

Thus, two asymptotic additive terms of the potential (\ref{Vcor}), 
$V_S(r)$ and $V_L(r)$, separately, yield the imaginary (\ref{En2Cou}) 
and real (\ref{E2Re}) momentum eigenvalues. These two asymptotic terms 
of the potential represent two ``different physics'' (coulombic OGE 
and linear string tension), therefore, two different realms of the 
interaction. Each of these two expressions, (\ref{En2Cou}) and 
(\ref{E2Re}), is exact and was obtained independently, therefore, we 
can consider the complex sum, $\pi_n={\rm Re}\,\pi_n+i{\rm Im}\,\pi_n$, 
given by eq. (\ref{PnCx}). Thus, we accept the complex momentum 
eigenvalues (\ref{PnCx}), that means the total energy and mass should 
be complex as well.\\

{\it 3.2. The complex mass}.

It is an experimental fact that the dependence $M_n^2(J)$ is linear 
for light mesons \ci{Collin}. However, at present, the best way to 
reproduce the experimental masses of particles is to rescale the 
entire spectrum given by (\ref{E2li}) assuming that the masses of 
the mesons are expressed by the relation \ci{Su}
\be 
M_n^2=E_n^2 -C^2, \label{Mn2C}\ee
where $C$ is a constant energy (shift parameter). Relation (\ref{Mn2C}) 
is used to shift the spectra and appears as a means to simulate the 
effects of unknown structure approximately. But, if we rewrite 
(\ref{Mn2C}) in the usual relativistic form,
\be 
M_n^2 =4\left[({\rm Re}\,\pi_n)^2+(\pm i\mu_I)^2\right], 
\label{MnImM} \ee
where Re$\,\pi_n$ is given by eq. (\ref{E2Re}), we come to the concept 
of the imaginary mass, $\mu_I$. Here in (\ref{MnImM}) we have introduced 
the notation, $4(\pm i\mu_I)^2=-C^2$. What is the mass $\mu_I$ and how 
to find it?

The Cornell potential (\ref{Vcor}) is usually written with the 
additive free parameter, $V_0$ \ci{Su,SeMPL}. In our approach the 
potential is a Lorentz-scalar and this parameter is included into the 
particle mass $m$ in the bound state, therefore, it does not appear 
apparently in final formulas. 

The required shift of the spectra naturally follows from the asymptotic 
solution of the semi-classical wave equation for the potential 
(\ref{Vcor}) \ci{SeMPL,SeEPL10}. To show that, we need to take into account 
the ``weak coupling effect'', i.e., together with the linear dependence 
in (\ref{E2Re}) we should include the contribution of the coulombic term, 
$-\alpha_s/r$, of the potential. That means, we need to solve the wave 
equation for the potential (\ref{Vcor}). 

These kind of calculations were done in \ci{SeMPL}. It was shown that 
for light mesons one may expect that the coulombic term can be 
considered as a small perturbation. Then we obtain the asymptotic 
expression similar to (\ref{Mn2C}) \ci{SeMPL,SeEPJC12},
\be
M_n^2=8\sigma(\tilde N -\tilde\alpha). \label{En2alf}
\ee
The additional term, $-8\tilde\alpha\sigma$, arises from the 
interference of the coulombic and linear components of the Cornell 
potential (\ref{Vcor}). Comparing (\ref{En2alf}) with (\ref{MnImM}), 
we obtain  
\be
\mu_I=\pm\sqrt{2\tilde\alpha\sigma}. \label{ImMas}\ee

The interference term $-8\tilde\alpha\sigma$ in (\ref{En2alf}) 
contains only the parameters of the potential (\ref{Vcor}) and is  
Lorentz-scalar, i.e., additive to the particle masses. This is why, 
we accept the last term in (\ref{En2alf}) to be the mass term, i.e., 
eq. (\ref{ImMas}) is the imaginary-part mass generated by the 
interference term of the Cornell potential (\ref{Vcor}). 

Thus, we have the particle real-part (constituent) mass, $\mu_R=m$, 
and the imaginary-part mass, $\mu_I$, originating from the potential. 
As in case of the eigenmomenta, we introduce the {\it complex} mass, 
\be 
\mu=m+i\sqrt{2\tilde\alpha\sigma}, \label{CompM}\ee 
which we use to reconstruct the complex masses of resonances, i.e., 
their centered masses and widths.\\

\centerline{\bf 4. The masses and widths of resonances}
\vskip 3mm\noindent
The complex-mass squared (\ref{E2nCx}) has the form of two free 
relativistic particles with the complex momenta and masses, 
\be 
\mc M_n^2=4\left[\left(\sqrt{2\sigma\tilde N}
-\frac{i\tilde\alpha m}N\right)^2
+(m+i\sqrt{2\tilde\alpha\sigma})^2\right]. \label{CompE2n}\ee
This expression can be written in another form as,
\be 
\mc M_n^2={\rm Re}\,\mc M_n^2+i{\rm Im}\,\mc M_n^2, 
\label{E2nReIm}\ee
where
\be 
{\rm Re}\,\mc M_n^2
=4\left({\rm Re}\,\pi_n^2+{\rm Re}\,\mu^2\right) 
\equiv 4\left[2\sigma\tilde N 
-\left(\frac{\tilde\alpha m}N\right)^2+m^2-\mu_I^2\right], 
\label{ReE2n}\ee 
\be 
{\rm Im}\,\mc M_n^2=8\left({\rm Re}\,\pi_n{\rm Im}\,\pi_n
+{\rm Re}\,\mu{\rm Im}\,\mu\right)
\equiv 8m\mu_I\left(1-\frac{\sqrt{\tilde\alpha\tilde N}}N\right). 
\label{ImE2n}\ee 
The real part (\ref{ReE2n}) exactly coincides with the interpolating 
mass formula (\ref{Mn2int}) obtained independently by another method 
\ci{SeEPJC12,SeEPL10,SeYF93,SeZC94}, and the imaginary one (\ref{ImE2n}) 
gives the resonance total width. 

In the pole approach, the parameters of resonances are defined in terms 
of the pole position $s_p$ in the complex $s$-plane as \ci{MorgPenn} 
\be
s_p=M_p^2-iM_p\Gamma_p, \label{Rpol}\ee
where $s=\mc M^2$ is the two-particle c.m. energy squared. 
Comparing (\ref{Rpol}) with eq. (\ref{E2nReIm}), we obtain the centered 
mass squared, $M_n^2$, given by eq. (\ref{ReE2n}), and the total width, 
\be 
\gn^{TOT}=-\frac{8m\mu_I}{M_n}
\left(1-\frac{\sqrt{\tilde\alpha\tilde N}}N\right). 
\label{Gamn}\ee

In general (mathematically), the S-matrix is a meromorphic function 
of complex variable $\mc M=\pm\sqrt s$, where the complex 
$s$-plane is replaced by the two-sheet Riemann surface, $\pm\sqrt s$, 
made up of two sheets $R_0$ and $R_1$, each cut along the positive 
real axis, Re$\mc M$, and with $R_1$ placed in front of $R_0$ 
\ci{Hislop,NieArri,FernOrtiz}. Square root of the complex expression 
(\ref{E2nReIm}) gives 
\be
(\mc M_n^2)^{1/2}=\pm\left[{\rm Re}\,\mc M_n+i\xi{\rm Im}\,\mc M_n\right], 
\label{SqCxMn}\ee 
where 
\be
{\rm Re}\,\mc M_n=\pm\sqrt{\frac{|\mc M_n^2|
+{\rm Re}\,\mc M_n^2}2}=M_n, \label{MnSq}\ee 
\be
{\rm Im}\,\mc M_n=\pm\sqrt{\frac{|\mc M_n^2|
-{\rm Re}\,\mc M_n^2}2}=-\frac{\gn^{TOT}}2. 
\label{GnSq}\ee 
Here $|\mc M_n^2|=\left[({\rm Re}\,\mc M_n^2)^2+
({\rm Im}\,\mc M_n^2)^2\right]^{1/2}$, 
$\xi={\rm sgn}({\rm Im}\,\mc M_n^2)$. 
The corresponding imaginary-part energy in agreement with (\ref{Mcx}) 
gives the total width of the resonance. 
Substituting (\ref{ReE2n}), (\ref{ImE2n}) in (\ref{MnSq}), 
(\ref{GnSq}), we obtain the process independent formulas for 
the centered masses and total widths of resonances.

Expressions (\ref{GnSq}) and (\ref{Gamn}) give close results for 
$\gn^{TOT}$. Imaginary-part mass (\ref{GnSq}) can be positive 
and negative, i.e., singularities are symmetrically located above and 
below of the real ${\rm Re}\,\mc M$-axis in the complex Riemann 
$\mc M$-surface. 

There is the lack of a precise definition of what is meant by mass 
and width of resonance. Comprehensive definitions of the centered 
mass and width of resonances require further investigations 
\ci{BerniCaPe}. Another definition of the resonance's width can be 
obtained from Eq. (\ref{ImE2n}). According to definition (\ref{MCx}) 
the width is given by the {\it imaginary-part} mass of the resonance's 
complex mass, $\mc M_n$. Dividing Eq. (\ref{ImE2n}) by $8m$ 
(exclusion of the real-mass term), we come to the following expression: 
\be 
\gn^{TOT}=\mu_I\left(1-\frac{\sqrt{\tilde\alpha\tilde N}}N\right). 
\label{GamnV}\ee
This width depends on the imaginary-part mass, $\mu_I$, of complex 
mass (\ref{MCx}) and quantum numbers. It has the maximum value 
given by the imaginary mass, $\mu_I$,
\be 
\gn^{max}=\sqrt{2\tilde\alpha\sigma}. \label{GamMax}\ee
Calculation results by these formulas are given below.

There exists a widely spread belief that the width of heavy hadronic 
resonances linearly depends on their mass. As one can see, in our 
analysis the total width of resonances $\gn^{TOT}$ is finite or 
decreasing; according to (\ref{GamMax}) it may have the maximum value, 
$\mu_I$. 

Complex poles of S-matrix always arise in conjugate pairs; 
poles in the left half-plane correspond to either bound or anti bound 
states \ci{LandLif,FernOrtiz}. If $\mc M_n=M_n-i\gn/2$ is 
a pole of S-matrix in the fourth quadrant of the surface $\pm\sqrt s$, 
then $\mc M_n=-M_n-i\gn/2$ is also a pole, but in the third 
quadrant (antiparticle) \ci{FernOrtiz}. Poles in the lower half-plane 
are complex-conjugated with zeros in upper half-plane. 

As an example, consider the $\rho$-family $J=l+1$ resonances which 
are located on the leading Regge trajectory, $\alpha_{\rho}(s)$. 
These resonances are the excitations of the $ud$ quark system. 
Calculation results for $M_n$ and total widths $\gn^{TOT}$ are 
given in Table 1, where masses and widths are in MeV.

\begin{table}[ht]
 \centering
 \begin{tabular}{llrrrcc}
\hline
\textrm{Meson}& $J^{PC}$ & $M_n^{ex}$ & $M_n^{th}$ & $\gn^{ex}$
& $\gn^{th}(\ref{GnSq})$ & $\gn^{th}(\ref{GamnV})$\\
\hline\hline
$\rho\ (1S)$&$1^{--}$&776 & 775 & 149 & 150 & 75\\
$a_2(1P)$ &$2^{++}$ &1318 & 1323 & 107 & 108 & 93\\
$\rho_3(1D)$&$3^{--}$& 1689 & 1689 & 161 & 170 & 188\\
$a_4(1F)$&$4^{++}$&1996 & 1985 & 255 & 194 & 249\\
$\rho_5(1G)$ &$5^{--}$& $-$ & 2234 & $-$ & 202 & 294\\
$a_6(1H)$&$6^{++}$& $-$& 2462 & $-$ & 205 & 328\\
\hline
 \end{tabular}
 \caption{Masses and total widths of the $\rho$-family resonances}
 \label{Tab1}
\end{table}

Parameter values in these calculations are found from the best fit 
to the available data: $\alpha_s=1.463$, $\sigma=0.134$\,GeV$^2$, 
$m=193$ MeV. Widths $\gn^{th}(\ref{GnSq})$ and 
$\gn^{th}(\ref{GamnV})$ are calculated with the use of eqs. 
(\ref{GnSq}) and (\ref{GamnV}). Maximum possible width 
$\gn^{max}$ according to (\ref{GamMax}) equals to the 
imaginary-part mass, $\gn^{max}=\mu_I=0.723$ GeV. Experimental 
data are from PDG-2012 \ci{PDG2012}.
\begin{figure}[ht]
 \centering
 \includegraphics[scale=1.1]{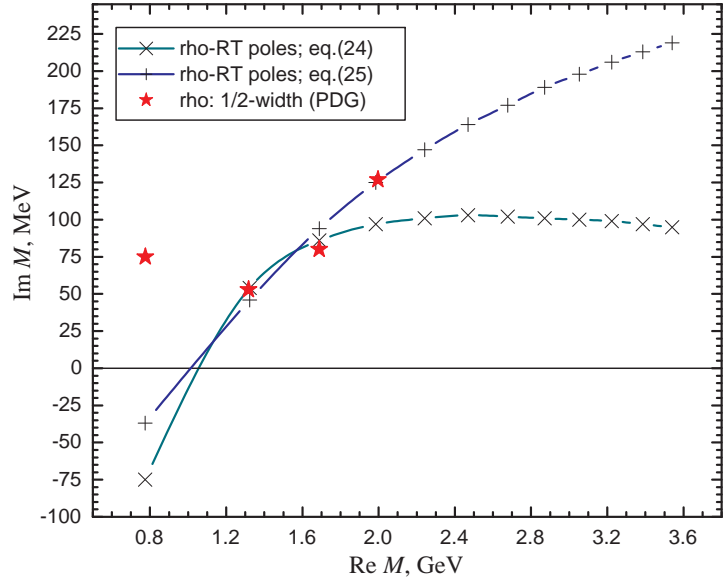}
 \caption{{\small The complex $\mc M$-surface. Stars show 
location of the complex-mass resonances for leading $\rho$ Regge 
trajectory, where the imaginary-part component is the resonance 
half-width; data are from PDG tables \protect\ci{PDG2012}. 
Crosses show calculations with the use of eqs. (\ref{GnSq}) and 
(\ref{GamnV}).}} \label{fig:CxMro}
\end{figure}
More accurate calculations require accounting for the spin 
corrections, i.e., spin-spin and spin-orbit interactions. The 
spin-dependent corrections to the potential (\ref{Vcor}) have been 
calculated in \ci{SeZC94} and can also be derived from lattice 
QCD, but we do not consider them here. In our case the spin corrections 
are accounted for effectively in parameters. 

Note one important feature of the $\rho$-family resonance data. 
There is a dip ($\gn^{TOT}=107$ MeV) for the $a_2(1320)$ 
resonance (see Fig. 1). This dip is described by our model and 
has the following explanation. According to eq. (\ref{GnSq}) 
the imaginary-part mass ${\rm Im}\,\mc M_n$ can be positive 
and negative, i.e., poles can be embedded in the first and 
fourth-quarter of the complex $\mc M$-surface. 
Imaginary-part mass ${\rm Im}\,\mc M_n$ of the $\rho(770)$ 
resonance is negative. That means that the corresponding pole 
is embedded in the fourth quadrant of the complex $\mc M$-surface 
(i.e., Re$\,\mc M>0$, Im$\,\mc M<0$). But other poles of the 
$\rho$-family resonances are located above the real axis, i.e., in 
the bound state region. 

\vskip 5mm{\bf 5. Conclusion}
\vskip 3mm
A thorough understanding of the physics summarized by the PDG is related 
to the concept of a resonance. Resonance in QM is closely connected with 
the concept of complex energy. In the version of QFT, the resonances are 
described by the complex-mass poles of the scattering matrix. 

In contrast to the usual analysis dealing with the scattering problem, 
we have studied mesonic resonances to be the quasi-bound eigenstates 
of two interacting quarks using the Cornell potential. 
Using the complex-mass scheme we have analyzed the exact eigenvalues 
for the coulombic and linear terms of the potential, separately, and 
obtained the complex-mass formula for meson resonances. 
This approach has allowed us to simultaneously describe in the unified 
way the masses and total widths of resonances in a good agreement with 
data.

The complex-mass expression (\ref{E2nCx}) may have relation to some 
non-hermitian Hamiltonian. 
In disagreement with a widely spread belief that the width of heavy 
hadronic resonances linearly depends on their mass, we have found this 
inconsistent with an existence of linear Regge trajectories. 
Widths obtained here are restricted or decreasing with the resonance 
mass.  

Resonances represent a very economical way in theoretical description 
of hadronic reactions at high energies. Such a task is very important 
nowadays since a great significance of the width of heavy resonances. 
Our analysis may be important for further development of the string 
model of hadrons and for improvement of such transport codes as the 
hadron string dynamics by including the finite width of heavy 
resonances. 

This work was done in the framework of investigations for 
the experiment ATLAS (LHC), code 02-0-1081-2009/2013, 
``Physical explorations at LHC'' (JINR-ATLAS).

\end{document}